\documentstyle[prb,multicol,aps,epsf]{revtex}

 \newcommand \be {\begin{equation}}
\newcommand \bea {\begin{eqnarray}}
\newcommand \ee {\end{equation}}
\newcommand \eea {\end{eqnarray}}
 \newcommand \eps {\epsilon}
 \newcommand \bi {\bibitem}

\newcommand \de {\delta}

\newcommand \th {\theta}
\newcommand \ph {\phi}

\newcommand \la {\lambda}

 \newcommand \al {\alpha}

\input{epsf}
\begin{document}
\draft      
\title{Solvable dynamics in a system of interacting random tops}
\author{Felix Ritort}
\address{Institute for Theoretical Physics\\ University of Amsterdam\\
Valckenierstraat 65\\ 1018 XE Amsterdam (The Netherlands).\\ E-Mail:
ritort@phys.uva.nl}

\date{\today}
\maketitle

\begin{abstract}
In this letter a new solvable model of synchronization dynamics is
introduced. It consists of a system of long range interacting tops with
random precession frequencies. The model allows for an explicit study of
orientational effects in synchronized phenomena. A stability analysis of
the incoherent solution is performed for different types of
orientational disorder. A system with only orientational disorder always
synchronizes in the absence of external noise.
\end{abstract} 




\vspace{.5cm}

One of the main goals in modern statistical physics concerns the
understanding of dynamical behavior in complex systems formed by a large
population of interacting entities. Much theoretical effort has been
devoted to the study of synchronization dynamics in simple systems
composed of interacting units. In those cases there is competition
between oscillations arising from the natural randomness in the members
of a population and macroscopic synchronization of the population as a
whole.  It is widely believed that these models provide a plausible
explanation for the existence of synchronization phenomena in a large
variety of physical systems ranging from biology to economy
\cite{WINFREE}. On the other hand, the study of this type of models
provides an interesting framework to study relaxational processes and
off-equilibrium behavior in physical systems driven by external
forces.

A simple model which describes the emergence of synchronization
phenomena in a population of phase oscillators was proposed many years
ago by Kuramoto \cite{KURA}. In this model each oscillator is
characterized by a random natural frequency and interacts
ferromagnetically with the rest by a mean-field interaction. The
ferromagnetic ordering of the system competes with the independent
random oscillation of each oscillator. Despite extensive studies in the
past, there are still some open issues such as a full characterization of
the stationary synchronized states, the study of the dynamics in the
absence of external noise \cite{SMM} as well as the onset of
synchronization in the critical region \cite{D,CR}.

The aim of this letter is to introduce a new model which yields insight
on the mechanisms involved in synchronization phenomena. The
fundamental new feature of this model is that it explicitly introduces
the role of orientational degrees of freedom in the synchronization
dynamics. This feature is not captured by the Kuramoto model and can be
interesting from the viewpoint of biological systems and also
neuroscience \cite{neuro} where orientational effects are certainly
important.

The model consists of a system of $N$ tops, each one characterized by a
random natural precession vector $\vec{\omega}_i$, interacting
ferromagnetically in a mean-field way. The tops are specified by a three
components unit vector $\vec{x}_i$ ($i=1,N$) and obey the following
dynamics,

\be
\frac{\partial \vec{x}_i}{\partial t}=-\frac{\partial{\cal H}}{\partial
\vec{x}_i}+\vec{\omega}_i \times \vec{x}_i\,+\vec{\eta}_i 
\label{eq1}
\ee

\noindent where the $\vec{\omega}_i$ are random quenched precession vectors,
${\cal H}$ is the Hamiltonian of the system and $\vec{\eta}_i$ is an
external white noise with zero mean and correlation
$\langle\vec{\eta}_i(t)\vec{\eta}_j(t')\rangle=6T\delta_{ij}
\delta(t-t')$. The Hamiltonian ${\cal H}$ is given by

\be
{\cal H}=-\frac{K}{N}\sum_{i<j} \vec{x}_i\vec{x}_j-\sum_i\mu_i\vec{x}_i^2
\label{eq1b}
\ee

where the $\mu_i$ are Lagrange multipliers introduced in order to ensure
the local constraint $|\vec{x}_i|=1$ at all times. Note that in the
dynamics there is competition between two opposite effects: a
ferromagnetic interaction which tries to align the tops in the same
direction and a natural precession of the tops around random quenched
directions which drives the system to an incoherent state. To analyze
the previous dynamics it is convenient to introduce polar coordinates
for the tops
$\vec{x}_i=(\sin\th_i\cos\ph_i,\sin\th_i\sin\ph_i,\cos\th_i)$ as
well as for the random precessions $\vec{\omega}_i=\omega_i
(\sin\mu_i \cos\la_i,\sin\mu_i\sin\la_i,\cos\mu_i)$. It is not difficult to
check that dynamical equation (\ref{eq1}) can be written in the
following way,

\bea
\frac{\partial\th_i}{\partial t}=-K(C\sin\th_i+
A\cos\th_i\cos\phi_i +B\cos\th_i\sin\phi_i)+\omega_i\sin\mu_i\sin(\la_i-\ph_i)+\al_i\label{eq2a}\\
\frac{\partial\ph_i}{\partial
t}\sin\th_i=-K (A
\sin\ph_i-B\cos\ph_i)\,+\,\omega_i(-\sin\mu_i\cos\th_i
\cos(\la_i-\ph_i)+\cos\mu_i\sin\th_i)+\beta_i
\label{eq2b}
\eea

\noindent where $NA=\sum_j\cos\ph_j\sin\th_j, NB=\sum_j\sin\ph_j\sin\th_j, 
NC=\sum_j\cos\th_j$, and the $\al_i,\beta_i$ are Gaussian noises given by,

\bea
\langle\al_i(t)\rangle=T\cot\th_i;~~~~~~~~\langle\al_i(t)\al_j(t')\rangle=
2T\de_{ij}\de(t-t')\\
\langle\beta_i(t)\rangle=0;~~~~~~~~ \langle\beta_i(t)\beta_j(t')\rangle=2T\de_{ij}\de(t-t')~~~~~~.
\eea

Note that the mean value of the noise $\al$ is not zero but depends on
the polar angle $\th$. This is expected because the angle $\th$ lies
always within an interval $(0,\pi)$ of length smaller than $2\pi$ and
$\th=0,\pi$ are returning points of the dynamics. Solving the previous
dynamical equations seems at first glance a rather difficult task. Here
we will follow a powerful approach recently introduced for the study of
the Kuramoto model \cite{PR} by considering an appropriate set of
moments which is invariant under the symmetry of the original dynamical
equations (\ref{eq1}).  Before studying the most general case, and for
sake of simplicity, we will first consider the case where there are no
frequencies, i.e. $\omega_i=0$.

{\em The non-disordered case $\omega_i=0$}. It is easy to observe from
equation (\ref{eq1}) that dynamics of the model is invariant under the
group of spatial rotations whose generators are $L^2$ and
$L^z$ and the eigenfunctions are the spherical harmonics. The most natural
set of moments which can close the dynamics is,

\be
M_{lm}(t)=\frac{1}{N}\sum_{i=1}^N Y_{lm}(\th_i(t),\ph_i(t))
\label{eq3}
\ee

\noindent where the $Y_{lm}(\th,\ph)=C_{lm}P_l^m(\th)\exp(im\ph)$
are the spherical harmonics with
$C_{lm}=((2l+1)(l-m)!/4\pi(l+m)!)^{\frac{1}{2}}$ as normalization
constants. $P_l^m$ are the associated Legendre polynomials where $-l\le
m\le l$.

It is not difficult to write the equation of motion for the $M_{lm}(t)$.
Using the definition of spherical harmonics, simple recursion formulas
for the Legendre polynomials as well as the Gaussian representation of
the noise, it is possible to show that the moments obey the following
closure equation,

\bea
\frac{\partial M_{l,m}}{\partial t}=K\bigl (C\,(a_{l,m}M_{l-1,m}-b_{l,m}M_{l+1,m})
-(A-iB)(c_{l,m}M_{l-1,m+1}+d_{l,m}M_{l+1,m+1})\nonumber\\
+(A+iB)(e_{l,m}M_{l-1,m-1}+f_{l,m}M_{l+1,m-1})\Bigr )-Tl(l+1)M_{l,m}
\label{eq4}
\eea

\noindent where the $a,b,c,d,e,f$ are Clebsch-Gordan-like coefficients
given by,

\bea
a_{l,m}=(l+1)\sqrt{\frac{l^2-m^2}{4l^2-1}};~~~~~~ b_{l,m}=
l\sqrt{\frac{(l+1)^2-m^2}{(2l+1)(2l+3)}}\label{eq6a}\\
c_{l,m}=\frac{(l+1)}{2}\sqrt{\frac{(l-m)(l-m-1)}{4l^2-1}};~~~~~~ d_{l,m}=
\frac{l}{2}\sqrt{\frac{(l+m+1)(l+m+2)}{(2l+1)(2l+3)}}\label{eq6b}
\eea

\noindent and $e_{l,m}=c_{l,-m},\,f_{l,m}=d_{l,-m}$. The time dependent
parameters $A,B,C$ in (\ref{eq4}) are given by
$A+iB=\sqrt{8\pi/3}M_{1,1}\,, C=\sqrt{4\pi/3}M_{1,0}$.  Note that the
dynamical equation (\ref{eq4}) is invariant under the transformation
$m\to -m$ if the moments satisfy the relation
$M_{l,-m}=(-1)^m\,M_{l,m}^*$ which is indeed the relation satisfied by
the spherical harmonics $Y_{l,m}$. The recursion relations
eq.(\ref{eq4}) show explicitly that the dynamics has been closed. The
special case $K=0$ corresponds to the random walk on a spherical
surface, a case well known in the literature \cite{LANG}. Then
eq.(\ref{eq4}) trivially reduces to $\dot M_{l,m}=-T l(l+1)M_{l,m}$
which shows that all moments decay exponentially fast to zero (except
$M_{0,0}$ which is a constant of motion equal to
$1/\sqrt{4\pi}$). Eq.(\ref{eq4}) can be expressed in a more intuitive
form introducing an appropriate generating function. To see this, let us
define

\be
g_t(\th,\ph)=\sum_{l=0}^{\infty}\sum_{m=-l}^lM_{l,m}(t)
Y_{l,m}(\th,\ph)=\frac{1}{N}\sum_{j=1}^{N}
\delta(\cos\th_j-\cos\th)\delta(\ph_j-\ph)
\label{eq7}
\ee

\noindent which is nothing else than the probability density to find a
top with a given solid angle $\Omega=(\th,\ph)$ on the unit sphere. The
last equality in eq.(\ref{eq7}) comes from the closure condition of the
spherical harmonics.  Using eq.(\ref{eq4}) it is not difficult to derive
the following equation for $g_t$: $\frac{\partial g_t}{\partial
t}=-div (g_t \vec{v}(\th,\ph))+ T\nabla^2 g_t$ where
$\vec{v}=(v_{\th},v_{\ph})$ is a two component velocity field given by,
$v_{\th}=Kr(\sin\Theta\cos\th\cos(\Phi-\ph)-\sin\th
\cos\Theta);~~v_{\ph}=Kr\sin\Theta\sin(\Phi-\ph)$. In polar coordinates
the equation for $g_t$ is:

\be
\frac{\partial g_t}{\partial
t}=\frac{1}{\sin\th}\Bigl (-\frac{\partial}{\partial\th}(g_t v_{\th}\sin\th)
-\frac{\partial}{\partial
\ph}(v_{\ph}g_t)+T(\frac{\partial}{\partial\th}(\sin\th\frac{\partial
g_t}{\partial\th})+\frac{1}{\sin\th}\frac{\partial^2 g_t}{\partial
\ph^2})\Bigr )~~~~~.
\label{eq8}
\ee

The time dependent parameters $r,\Theta,\Phi$ which appear in the
velocity field are self-consistently
computed from the probability distribution $g_t$. They are given by
$A=r\sin(\Theta)\cos(\Phi); B=r\sin(\Theta)\sin(\Phi); C=\cos(\Theta)$
where $A,B,C$ have been introduced before. $r=\sqrt{A^2+B^2+C^2}$ is the
synchronization parameter and measures the degree of coherence of the
tops. Equation (\ref{eq8}) seems hardly manageable but still some
results can be inferred, in particular the nature of the stationary
solutions. It is easy to check that these are solutions of the Boltzmann
type, i.e.  equilibrium solutions of the Hamiltonian
eq.(\ref{eq1}). This is expected since the model, in the absence of
random precessions, is purely relaxational. We will not extend further
on the connections between the moment approach shown here and the
probability density formalism. The equation (\ref{eq8}) is formally
identical to that derived for the Kuramoto model with the only
difference that its velocity field has now two components. Certainly it
can be generalized for O(n) vector mean-field models with any number of
components (n=2 is the Kuramoto model and n=3 the present case). Since
our main interest in this note is to analyze the model with random
precession frequencies, we will come back to the moment approach which
is more appropriate and suited to investigate the dynamical phase
diagram of the disordered model.

{\em The disordered case $\omega_i\ne 0$}. In the presence of random
precessions the model is not purely relaxational since there are
external driving random forces. In this case we expect the emergence of
a rich dynamical behavior due to the competition between the ordering
ferromagnetic interaction and the random natural precessions of the
tops. For the sake of simplicity and in order to investigate the effect of
orientational disorder, we will consider here the case in which
the precession angular velocities have the same magnitude
(i.e. $\omega_i=\omega$), but point in different directions in space. In
this case the disorder is specified by a probability distribution
$p(\mu,\la)$. It is easy to generalize the definition of the moments
(\ref{eq3}) to include the presence of quenched disorder. Now the
moments are characterized by four quantum numbers \cite{com1}, two of
them appearing as a consequence of the disorder,

\be
M_{l,m}^{p,q}=\frac{1}{N}\sum_{i=1}^N\,Y_{l,m}(\th_i,\ph_i)Y_{p,q}(\mu_i,\la_i)~~~~.
\label{eq9}
\ee

After some algebra, the closure equations for the new set of moments read

\bea \frac{\partial M_{l,m}^{p,q}}{\partial t}=K\bigl
(C\,(a_{lm}M_{l-1,m}^{p,q}-b_{l,m}M_{l+1,m}^{p,q})
-(A-iB)(c_{l,m}M_{l-1,m+1}^{p,q}+d_{l,m}M_{l+1,m+1}^{p,q})
+\nonumber\\
(A+iB)(e_{l,m}M_{l-1,m-1}^{p,q}+f_{l,m}M_{l+1,m-1}^{p,q})\Bigr )-
Tl(l+1)M_{l,m}^{p,q}+
\frac{i\omega}{2}(A_{l,m}^{p,q}M_{l,m}^{p-1,q+1}\label{eq10}\\
-B_{l,m}^{p,q}M_{l,m-1}^{p+1,q+1}-C_{l,m}^{p,q}M_{l,m+1}^{p-1,q-1}+
D_{l,m}^{p,q}M_{l,m+1}^{p+1,q-1})+
im\omega(E_{p,q}M_{l,m}^{p+1,q}+F_{p,q}M_{l,m}^{p-1,q})\nonumber \eea

where the expressions for $a,b,c,d,e,f$ have been given in
eq.(\ref{eq6a},\ref{eq6b}) and the other coefficients are

\bea
A_{l,m}^{p,q}=\sqrt{\frac{(l-m+1)(l+m)(p-q)(p-q-1)}{4p^2-1}};~~ 
E_{p,q}=\sqrt{\frac{(p+q+1)(p-q+1)}{(2p+1)(2p+3)}}\label{eq11a}\\
B_{l,m}^{p,q}=\sqrt{\frac{(l-m+1)(l+m)(p+q+2)(p+q+1)}{(2p+1)(2p+3)}};~~
F_{p,q}=\sqrt{\frac{(p-q)(p+q)}{4p^2-1}}\label{eq11b}
\eea

\noindent and $C_{l,m}^{p,q}=A_{l,-m}^{p,-q},\,
D_{l,m}^{p,q}=B_{l,-m}^{p,-q}$. The parameters $A,B,C$ have been already
defined after eqs.(\ref{eq2a},\ref{eq2b}). In terms of the new set of
moments they are given by, $A+iB=4\pi\sqrt{2/3}M_{1,1}^{0,0};
C=4\pi\sqrt{1/3}M_{1,0}^{0,0}$. Note that the set of moments
$M_{0,0}^{p,q}= \frac{1}{\sqrt{4\pi}}\int d\Omega Y_{p,q}(\Omega)
p(\Omega)$ (where $\Omega=(\mu,\la)$ is the solid angle and the
integration is over all the unit sphere) are constants of the motion.
Let us mention that also in this case a probability distribution, i.e. a
generating function for all the moments, can be defined like in the
non-disordered case. We will not extend on these considerations and
instead we will focus on the novel properties of the model. For this
purpose let us consider the case of disorder distributions with axial
symmetry around the z-axis (the following considerations can be easily
extended to the more general case \cite{Rit}). This means that
$p(\Omega)=p(\mu)$. In this case it is possible to study the region in
the plane $\tilde{\omega}=\omega/T$,$\tilde{K}=K/T$ where the incoherent
solution $r=0$ is unstable. Our calculation follows essentially the
equivalent one for the Kuramoto model \cite{STMI,BNS}. To this end we
expand the moments around the incoherent solution
$M_{l,m}^{p,q}=\frac{1}{\sqrt{4\pi}}\delta_{l,0}\delta_{m,0}M_{0,0}^{p,q}
+ \eps\eta_{l,m}^{p,q}\exp(\alpha t)$ where $\eps$ is a small parameter.
Substituting this result in equation (\ref{eq10}) and performing a
linear stability analysis we find a set of two linear equations
(uncoupled from the rest of modes) involving the two fundamental modes
$\eta_{1,0}(\Omega),\eta_{1,1}(\Omega)$. These modes are defined by,
$\eta_{l,m}^{p,q}=\int d\Omega \eta_{l,m}(\Omega) Y_{p,q}(\Omega)$.
After some calculations (details will be shown elsewhere \cite{Rit}) the
condition for the linear stability of the incoherent solution is
determined by the roots of the equation $\la_1(\gamma)\la_2(\gamma)=0$
where,

\bea
\la_1(\gamma)=\gamma^3-\frac{2\tilde{K}}{3}\gamma^2+\tilde{\omega}^2\gamma-
\frac{2\tilde{K}}{3}\tilde{\omega}^2\overline{\cos^2(\mu)}\label{eq12a}\\
\la_2(\gamma)=\gamma^3-\frac{2\tilde{K}}{3}\gamma^2+\tilde{\omega}^2\gamma-
\frac{\tilde{K}}{3}\tilde{\omega}^2\overline{\sin^2(\mu)}~~~,~~~\gamma=\alpha+2
\label{eq12b}
\eea

\noindent and
$\overline{A(\mu)}=2\pi\int_0^{\pi}\sin(\mu)p(\mu)A(\mu)$. These are two
cubic equations and each one yields three roots (one of them is real and
the other two are complex conjugates). When the real part of one of
these six roots becomes larger than $2$ the incoherent solution is
unstable. The equation in (\ref{eq12a},\ref{eq12b}) that determines the
stability is the one where the independent $\gamma$ term is the largest
(see the examples below). The boundaries of the region where the
incoherent solution is unstable indicate the dynamical transition
lines. While the stability boundaries will depend on the particular
disorder distribution, some general results can still be inferred. In
particular, in the absence of noise $T=0$, there is no critical value of
$K$ below which the incoherent solution becomes linearly stable. This
result is also found in the Kuramoto model with disorder distributions
with vanishing probability of oscillators with zero frequency
\cite{STMI}, for instance, the bimodal distribution \cite{BNS}.  This
last case has been shown to display a very rich dynamical behavior also
shared by the present model. Here we are going to consider three cases
of disorder distributions which cover a large variety of physical
situations: a) precession vectors lying in the z-axis and randomly
pointing in opposite directions, i.e. $p(\mu)=0$ except for $\mu=0,\pi$;
b) precession vectors uniformly distributed and lying in the x-y plane,
i.e.  $p(\mu)=\frac{1}{2\pi}\delta(\mu-\frac{\pi}{2})$; c) precession
vectors isotropically distributed on the sphere,
i.e. $p(\mu)=\frac{1}{4\pi}$.

Case a) is particularly interesting despite its simplicity. Half of the
tops precess in one sense, the other half do in the other sense, around
the z-axis. In this case $\overline{\cos^2(\mu)}=1$ and the stability
condition for the incoherent solution is determined by $\la_1(\gamma)$
which yields $\tilde{K}<3$. For $\tilde{K}>3$ the incoherent solution is
always unstable whatever the value of $\tilde{\omega}$. A quick
inspection on the definition of the moments eq.(\ref{eq9}) reveals that
they reduce to two families $H_{l,m}=M_{l,m}^{0,0},
G_{l,m}=M_{l,m}^{1,0}$. The dynamical equations for the moments $H$ and
$G$ are very similar to those obtained for the Kuramoto model in case of
a bimodal distribution \cite{PR}, the crucial difference being that the
two families of moments $H$ and $G$ are coupled each other with a term
proportional to the quantum number $m$. A trivial stationary solution of
dynamical equations is then given by $H_{l,m}=H_{l,0}\delta_{m,0},
G_{l,m}=0$. This solution coincides with that derived previously in the
non-disordered case $\omega_i=0$. Interestingly enough we find that
the stationary solutions of the model (a) are the equilibrium solutions
of the Hamiltonian eq.(\ref{eq1b}) even though there is no effective
Hamiltonian which yields the original equation of motion (\ref{eq1})
\cite{WH}.

Cases b) and c) display a more rich dynamical behavior. Case b)
corresponds to all precession directions lying on the x-y plane. In this
case $\overline{\sin^2(\mu)}=1$ and the stability criteria is fixed by
the roots of $\la_2(\gamma)$. The incoherent solution is linearly stable
for $\tilde{K}<3$, linearly unstable for $\tilde{K}>6$ and for
intermediate values of $\tilde{K}$ depending on the value of
$\tilde{\omega}$. More concretely, the incoherent solution is stable for
$\tilde{\omega}^2>(8\tilde{K}-24)/(6-\tilde{K})$.  Case c) is the limit
situation (i.e. the most orientationally disordered case) with the
largest region in the phase diagram where the incoherent solution is
linearly stable.  In this case, $\overline{\cos^2(\mu)}=1/3$ and
$\la_1(\gamma)=\la_2(\gamma)$.  The incoherent solution is linearly
stable for $\tilde{K}<3$, linearly unstable for $\tilde{K}>9$ and stable
in the intermediate region $3<\tilde{K}<9$ if
$\tilde{\omega}^2>(12\tilde{K}-36)/(9-\tilde{K})$.


The main difference between the present model and the Kuramoto model
with bimodal distribution of frequencies lies in their phase
diagrams. The neutral stability line for the incoherent solution in the
phase diagram $(\tilde{K},\tilde{\omega})$ has a vertical parabolic
branch at $\tilde{K}_c$: $\tilde{\omega}\sim
(\tilde{K}-\tilde{K_c})^{-\frac{1}{2}}$ ($\tilde{K}_c=6,9$ for cases b)
and c) respectively). In the Kuramoto model, the neutral stability line
is the semirect $\tilde{K}_c=4,\tilde{\omega}> 1$ \cite{BNS}.  In the
region where the incoherent solution is unstable we expect the existence
of coherent fixed point solutions as well as oscillatory time dependent
solutions. To show this we have numerically solved eq.(\ref{eq10}) for a
finite set of moments using a simple second order Euler algorithm. A
maximum value of both angular momenta $l$ an $p$ equal to 10 is enough
for the calculation (which implies an approximate number of 10000
moments). In figure 1 we show the time evolution of the synchronization
parameter for model c) in the region where the incoherent solution is
unstable starting from two different initial conditions.

\begin{figure}
\centerline{\epsfxsize=8cm\epsffile{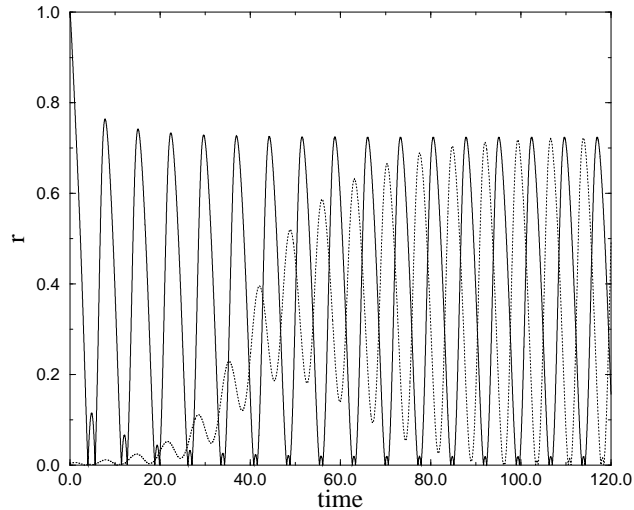}}

\caption{Synchronization parameter $r$ as a function of time for case c)
at $\tilde{K}=\tilde{\omega}=50/3$ starting from two initial 
conditions: all tops pointing in the same directions (continuous line)
and the uncoherent state (dotted line).}
\end{figure}

To summarize, we have introduced an solvable model of interacting random
tops. The model explicitly introduces orientational effects in the
synchronization dynamics, a feature which is not present in the Kuramoto
model. The model has been analytically studied in the orientationally
disordered case introducing a family of moments (depending on four
quantum numbers, two of them arising from the randomness in the system)
which can be exactly closed due to the symmetry invariances of the
original dynamical equations (\ref{eq1}). The moment formalism is simple
and suited for theoretical analysis and this has been explicitly checked
in the study of the stability properties of the incoherent solution. We
find that the system with only orientational disorder always
synchronizes in the absence of external noise. The present model (n=3)
altogether with the Kuramoto model (n=2) provide simple examples of
analytically tractable models with non relaxational dynamics. There
still remain some open issues like the study of the effect of frequency
dispersion (i.e. the $\omega_i$ different) as well as the noise free
($T=0$) dynamics in this model where the higher modes of the incoherent
solution are neutrally stable.

{\bf Acknowledgments.}  F.R is grateful to the Foundation for
Fundamental Research of Matter (FOM) in The Netherlands for financial
support through contract number FOM-67596. I acknowledge
I. Pagonabarraga, C. J. Perez-Vicente and J. M. Rub\'{\i} for useful
discussions and L. L. Bonilla for a careful reading of the manuscript.


\begin{thebibliography}{99}

\bibitem{WINFREE} A.T. Winfree, {\em The geometry of biological time}
Springer Verlag (1980).

\bibitem{KURA}  Y. Kuramoto, International Symposium on Mathematical
Problems in Theoretical Physics. Lectures notes in Physics {\bf 39}, 420
(Springer, Berlin) (1975)

\bi{SMM} S. H. Strogatz, R. E. Mirollo and P. C Matthews,
Phys. Rev. Lett. {\bf 68}, 2730 (1992);

\bi{D} H. Daido, Phys. Rev. Lett. {\bf 73}, 760 (1994);
 
\bi{CR} J. D. Crawford, Phys. Rev. Lett. {\bf 74}, 4341 (1995).
 
\bi{neuro} C. M. Gray, P. Konig, A. K. Engel and W. Singer, Nature {\bf
338}, 334 (1989)

\bi{PR} C. J. Perez-Vicente and F. Ritort, {\em A moment based approach
to the dynamical solution of the Kuramoto model}, submitted to J. Phys. A
(lett).

\bi{com1} We will denote the indices of the moments as quantum numbers in
analogy with the labelling of eigenstates in quantum mechanics.

\bi{LANG} W. T. Coffey, Y. P. Kalmykov and J. T. Waldron, {\em The
Langevin equation}, World Scientific (1996)

\bi{Rit} F. Ritort, to be published

\bi{STMI} S.H. Strogatz, R.E. Mirollo, J. Phys. A. Math. Gen. {\bf 21},
699 (1988); J.  Stat. Phys. {\bf 63}, 613 (1991)

\bi{BNS} L.L. Bonilla, J.C. Neu and R. Spigler, J. Stat. Phys. 
{\bf 67},313 (1992). 

\bi{WH} In the Kuramoto model there have been attempts to define a Lyapunov
function for the dynamics, see for instance 
J.L. Van Hemmen and W.F. Wrezinsky, J. Stat. Phys. {\bf72}, 145 (1993).

\end{thebibliography}
\end{document}